\documentclass[11pt,a4paper]{article}
\usepackage[english]{babel}
\usepackage{amsmath,amsthm,amssymb,epsfig,latexsym}
\usepackage[numbers,square,sort&compress]{natbib}
\usepackage{color}

\newcommand{\la}{u}

\newcommand{\bucb}{\bar{u}^{\scriptscriptstyle C,B}}

\newcommand{\bvcb}{\bar{v}^{\scriptscriptstyle C,B}}
\newcommand{\so}{\scriptscriptstyle \rm I}
\newcommand{\st}{\scriptscriptstyle \rm I\hspace{-1pt}I}

\newcommand{\B}{\mathbb{B}}
\newcommand{\C}{\mathbb{C}}

\newcommand{\bu}{\bar u}
\newcommand{\bv}{\bar v}
\newcommand{\bub}{\bar u^{\scriptscriptstyle B}}
\newcommand{\buc}{\bar u^{\scriptscriptstyle C}}
\newcommand{\bvb}{\bar v^{\scriptscriptstyle B}}
\newcommand{\bvc}{\bar v^{\scriptscriptstyle C}}

\newcommand{\be}[1]{\begin{equation}\label{#1}}
\newcommand{\ba}[1]{\begin{multline}\label{#1}}
\newcommand{\ee}{\end{equation}}
\newcommand{\ea}{\end{eqnarray}}

\newcommand{\num}{\\\rule{0pt}{20pt}}

\newcommand{\diag}{\mathop{\rm diag}}

\newcommand{\str}{\mathop{\rm str}}

\newtheorem{thm}{Theorem}[section]
\newtheorem{prop}{Proposition}[section]

 \makeatletter
 \@addtoreset{equation}{section}
 \makeatother
 
\begin{document}

\vspace{12pt}

\begin{center}
\begin{LARGE}
{\bf Form factors of local operators\\[1.ex]
in supersymmetric quantum integrable models }
\end{LARGE}

\vspace{40pt}

\begin{large}
{J.~Fuksa${}^{a,b}$,  N.~A.~Slavnov${}^c$\footnote{fuksa@theor.jinr.ru, nslavnov@mi.ras.ru}}
\end{large}

 \vspace{12mm}

\vspace{4mm}

${}^a$ {\it BLTP, Joint Institute of Nuclear Research, Dubna,  Russia}

\vspace{4mm}

${}^b$ {\it FNSPE, Czech Technical University in Prague, Czech Republic}

\vspace{4mm}

${}^c$ {\it Steklov Mathematical Institute of Russian Academy of Sciences,
Moscow, Russia}

\end{center}


\vspace{4mm}


\begin{abstract}
We apply  the nested algebraic Bethe ansatz to the models with $\mathfrak{gl}(2|1)$ and $\mathfrak{gl}(1|2)$ supersymmetry.
We show that form factors of local operators in these models can be expressed in terms of the universal form factors.
Our derivation is based on the use of the $RTT$-algebra only. It does not refer to any specific representation of this algebra.
We obtain thus determinant representations for form factors of local operators in the cases where an explicit solution of the
quantum inverse scattering problem is not known.
\end{abstract}

\vspace{1cm}

\vspace{2mm}

\section{Introduction}

Quantum Inverse Scattering Method (QISM) is a powerful tool for solving quantum integrable models \cite{TakF79,SklTF79,KorBI93L,FadLH96}. This method
allows one to find spectra of quantum Hamiltonians  via the algebraic Bethe ansatz. The main advantage of the
algebraic Bethe ansatz is that it gives a systematic procedure to describe the spectra of the models, which might have
completely different physical interpretation. This is because this method only deals with the algebra of the monodromy
matrix entries, but not with its specific representation.

The QISM and the algebraic Bethe ansatz also can be used for calculation of form factors and correlation functions \cite{IzeK84,KorBI93L,KitMST05,Sla07}.
Similarly to the problem
of the Hamiltonian spectrum, in many cases this method gives quite general results, which can be used for the study of a wide class of models.
In particular, it was shown recently \cite{PakRS15d,PakRS15b} that form factors of local operators (FFLO) in the models with $\mathfrak{gl}(3)$-invariant $R$-matrix are all expressed in terms of {\it universal form factors} \cite{PakRS15a}. The latter are completely determined by
the $R$-matrix and do not depend on the model under consideration.

Knowing FFLO one can solve the problem of correlation functions via their form factor expansion. In the models,
for which an explicit solution of the quantum inverse scattering problem is known \cite{KitMT99,MaiT00}, the FFLO
are directly related to the ones of the monodromy matrix entries. Using this result, correlation functions of the $XXZ$ spin chain
and other integrable models were studied in the
series of works \cite{KitMST05,KitKMST11,KitKMST12,CauM05,CauPS07}. It is worth mentioning, however, that the existence of an  explicit solution of the
quantum inverse scattering problem is based on a specific representation of the underlying $RTT$-algebra. It can be found for various spin chains, but not
in general.

In the present paper we consider models described by the $\mathfrak{gl}(2|1)$ and $\mathfrak{gl}(1|2)$ superalgebras. Determinant representations for form factors of the monodromy matrix entries in these models were found in \cite{HutLPRS16a}. These formulas can be directly applied to the calculation of
the FFLO in the  supersymmetric t-J model \cite{Sch87,For89,EssK92,FoeK93,Gom02}, because  the quantum
inverse scattering problem for this model was solved in \cite{GohK00}.
Our goal is to generalize these results. Namely, we show that similarly to the $\mathfrak{gl}(N)$ case, the FFLO in the
models with $\mathfrak{gl}(2|1)$ and $\mathfrak{gl}(1|2)$ supersymmetries are proportional to the universal form factors.

Our method is based on the composite model \cite{IzeK84}. It was used in \cite{PakRS15b,PakRS15c} for the calculation of the FFLO
in the $\mathfrak{gl}(3)$-based models. In the present paper we generalize this approach to the case of superalgebras. Due to an isomorphism
between Yangians of $\mathfrak{gl}(2|1)$ and $\mathfrak{gl}(1|2)$ \cite{PakRS17} we consider below the case of $Y\bigl(\mathfrak{gl}(2|1)\bigr)$ only.
It is worth mentioning that our method does not use any specific representation of the $RTT$-algebra. Therefore, the obtained results have the same degree of generality as the Bethe equations which are used to determine the spectra of the quantum Hamiltonians.

The paper is organized as follows. In section~\ref{S-BNN} we introduce the model under consideration and describe the basic
notions and the notation used in the paper. In section~\ref{SS-CM} we recall the notion of the composite model and introduce partial
zero modes of the monodromy matrix. There we explain how the use of the partial
zero modes allows one to relate different FFLO with each other. The main results of the paper are collected in
section~\ref{S-MR}. In section~\ref{S-FFOD} we prove the results of section~\ref{S-MR}. A part of the proof is placed in appendix~\ref{A-CALC}.

\section{Basic notions and notation\label{S-BNN}}

In this section we briefly describe basic notions of $\mathfrak{gl}(2|1)$-based integrable models solvable by the algebraic Bethe ansatz.
The reader can find a more detailed description in \cite{KulS80,PakRS17,BeRa08,HutLPRS17a}.

\subsection{Graded models and Bethe vectors\label{SS-GMBV}}

The central object of the algebraic Bethe ansatz method is a quantum monodromy matrix $T(u)$. It
acts in the tensor product  $V\otimes \mathcal{H}$, where $\mathcal{H}$ is a Hilbert space of a quantum model and
$V$ is an auxiliary space. Commutation relations between the entries $T_{ij} (u)$ are gathered in
an $RTT$-relation
\begin{equation}\label{L-op-com}
R(u,v)\cdot (T(u)\otimes \mathbb{I})\cdot (\mathbb{I}\otimes T(v))=
(\mathbb{I}\otimes T(v))\cdot (T(u)\otimes \mathbb{I})\cdot R(u,v).
\end{equation}
Here $R(u,v)$ is an $R$-matrix acting in the tensor product $V\otimes V$ of the auxiliary vector spaces $V$.  Equation \eqref{L-op-com} holds in the tensor
product $V\otimes V\otimes\mathcal{H}$.

For $\mathfrak{gl}(2|1)$-based models the auxiliary vector space $V$ is a ${\mathbb Z}_2$-graded space ${\mathbb C}^{2|1}$  with a basis $\{{\rm e}_1,
{\rm e}_2,{\rm e}_3\}$. We call the vectors  $\{{\rm e}_1,{\rm e}_2\}$  even, while ${\rm e}_{3}$ is odd. Respectively, we  introduce  a parity function on the set of indices as  $[1]=[2]=0$  and $[3]=1$.

The $R$-matrix in \eqref{L-op-com} has the form
\begin{equation}\label{DYglmn}
R(u,v)\ =\  \mathbb{I}+g(u,v)\mathbb{P}, \qquad g(u,v)=\frac{c}{u-v},
\end{equation}
where $c$ is a constant, $\mathbb{I}$ is the identity matrix in $V\otimes V$, and $\mathbb{P}$ is the graded 
permutation matrix \cite{KulS80}. The tensor product in  \eqref{L-op-com}
 is also graded leading to the set of commutation relations between the monodromy matrix entries $T_{ij}$:
\begin{equation}\label{TM-1}
[T_{ij}(u),T_{kl}(v)\}
=(-1)^{[i]([k]+[l])+[k][l]}g(u,v)\Big(T_{kj}(v)T_{il}(u)-T_{kj}(u)T_{il}(v)\Big),
\end{equation}
where we have introduced a graded commutator as
\be{Def-SupC}
[T_{ij}(u),T_{kl}(v)\}= T_{ij}(u)T_{kl}(v) -(-1)^{([i]+[j])([k]+[l])}   T_{kl}(v)  T_{ij}(u).
\ee
The supertrace of the monodromy matrix
 \be{Strace}
 \mathcal{T}(u)=\str T(u)=\sum_{i=1}^3 (-1)^{[i]}T_{ii}(u)
 \ee
 is called the transfer matrix. It is a generating function of the integrals of motion of the integrable model under consideration.
 The transfer matrix eigenstates are called {\it on-shell Bethe vectors}. They play an important role in the considerations below.

We assume that the space $\mathcal{H}$, in which the operators $T_{ij}$ act, contains a pseudovacuum vector $|0\rangle$. This vector possesses the following
properties:
\be{vac}
\begin{aligned}
T_{ii}(u)|0\rangle&=\lambda_i(u)|0\rangle,\\
T_{ij}(u)|0\rangle&=0,\qquad i>j,
\end{aligned}
\ee
where $\lambda_i(u)$ are some functions of complex variable $u$. A specific choice of these functions means fixing of a specific
integrable model. For us, however, they remain free functional parameters. This treatment of $\lambda_i(u)$ allows us
to consider a wide class of integrable models within a common framework.

We also assume that the operators $T_{ij}$  act in the dual space $\mathcal{H}^*$ with a dual pseudovacuum vector $\langle0|$. This vector
has analogous properties
\be{dvac}
\begin{aligned}
\langle0|T_{ii}(u)&=\lambda_i(u)\langle0|,\\
\langle0|T_{ij}(u)&=0,\qquad i<j,
\end{aligned}
\ee
where $\lambda_i(u)$ are the same as in \eqref{vac}. Below we use the ratios of these functions
 \be{ratios}
 r_1(w)=\frac{\lambda_1(w)}{\lambda_2(w)}, \qquad  r_3(w)=\frac{\lambda_3(w)}{\lambda_2(w)}.
 \ee

Bethe vectors of $\mathfrak{gl}(2|1)$-invariant models are certain polynomials in operators $T_{ij}(u)$ with $i<j$ acting on the pseudovacuum vector. Their
explicit form was found in \cite{PakRS17} (see also \cite{HutLPRS17a} for the general $\mathfrak{gl}(m|n)$ case). They depend on two sets of variables called Bethe parameters.
We denote the Bethe vectors $\mathbb{B}_{a,b}(\bu;\bv)$. Here the Bethe parameters are $\bu=\{u_1,\dots,u_a\}$ and
$\bv=\{v_1,\dots,v_b\}$. The subscripts $a$ and $b$ ($a,b=0,1,\dots$) denote the cardinalities of the sets $\bu$ and $\bv$ respectively.

Similarly one can construct dual Bethe vectors in the dual space $\mathcal{H}^*$ as polynomials in operators $T_{ij}(u)$ with $i>j$ acting on the
dual pseudovacuum vector $\langle0|$ \cite{PakRS17}. We denote them $\mathbb{C}_{a,b}(\bu;\bv)$ with the same meaning of the arguments and
the subscripts.

For generic (dual) Bethe vectors the Bethe parameters $\bu$ and $\bv$ are generic complex numbers. If these parameters satisfy a system of Bethe equations
\be{BE}
\begin{aligned}
r_1(u_j)&=\prod_{\substack{k=1\\k\ne j}}^a\frac{f(u_j,u_k)}{f(u_k,u_j)}\prod_{l=1}^bf(v_l,u_j),\qquad j=1,\dots,a,\\
r_3(v_j)&=\prod_{l=1}^af(v_j,u_l),\qquad j=1,\dots,b,
\end{aligned}
\ee
where
\be{f}
f(u,v)=1+g(u,v)=\frac{u-v+c}{u-v},
\ee
then the corresponding (dual) Bethe vector becomes on-shell. It means that this is an eigenvector of the transfer matrix
$ \mathcal{T}(u)$ \eqref{Strace}

\be{Left-act}
\mathcal{T}(w)\mathbb{B}_{a,b}(\bu;\bv)= \tau(w|\bu,\bv)\,\mathbb{B}_{a,b}(\bu;\bv),\qquad
 \mathbb{C}_{a,b}(\bu;\bv)\mathcal{T}(w) = \tau(w|\bu,\bv)\,\mathbb{C}_{a,b}(\bu;\bv),
\ee
where the eigenvalue $\tau(w|\bu,\bv)$ is
\be{tau-def}
\tau(w|\bu,\bv)=\lambda_1(w)\prod_{j=1}^af(u_j,w)+ \lambda_2(w)\prod_{j=1}^af(w,u_j)\prod_{k=1}^b f(v_k,w)
 -\lambda_3(w)\prod_{k=1}^bf(v_k,w).
\ee

Besides the monodromy matrix $T(u)$ we also consider a twisted monodromy matrix $T_\kappa(u)=\kappa T(u)$, where $\kappa$ is a
diagonal matrix $\kappa=\diag\{\kappa_1,\kappa_2,\kappa_3\}$, and $\kappa_i$ are complex numbers. The supertrace
$\mathcal{T}_\kappa(u)=\str T_\kappa(u)$ is called the twisted transfer matrix. A generic Bethe vector becomes an eigenstate of the
twisted transfer matrix, if the Bethe parameters satisfy a system of twisted Bethe equations
\be{TBE}
\begin{aligned}
r_1(u_j)&=\frac{\kappa_2}{\kappa_1}\prod_{\substack{k=1\\k\ne j}}^a\frac{f(u_j,u_k)}{f(u_k,u_j)}\prod_{l=1}^bf(v_l,u_j),\qquad j=1,\dots,a,\\
r_3(v_j)&=\frac{\kappa_2}{\kappa_3}\prod_{l=1}^af(v_j,u_l),\qquad j=1,\dots,b.
\end{aligned}
\ee
The corresponding (dual) Bethe vector is then called the twisted (dual)  on-shell Bethe vector. The twisted transfer matrix eigenvalue
on the vector $\mathbb{B}_{a,b}(\bu;\bv)$ is given by
\be{ktau-def}
\tau_\kappa(w|\bu,\bv)=\kappa_1\lambda_1(w)\prod_{j=1}^af(u_j,w)+ \kappa_2\lambda_2(w)\prod_{j=1}^af(w,u_j)\prod_{k=1}^b f(v_k,w)
 -\kappa_3\lambda_3(w)\prod_{k=1}^bf(v_k,w).
\ee
The use of the twisted monodromy matrix and the twisted on-shell Bethe vectors allows us to construct a special generating functional for
FFLO (see section~\ref{S-FFOD}).

\subsection{Shorthand notation\label{SS-SN}}

We denote sets of variables by bar:  $\bu$, $\bv$ etc. If necessary, the cardinalities of the sets are
given in special comments.
Individual elements of the sets are denoted by latin subscripts: $u_j$, $v_k$ etc.
We say that $\bar x=\bar x'$,
if $\#\bar x=\#\bar x'$ and $x_i=x'_i$ (up to a permutation) for $i=1,\dots,\#\bar x$. We say that $\bar x\ne \bar x'$ otherwise.

Below we consider partitions of the sets into subsets.  The notation $\bar u\Rightarrow\{\bu_{\so},\bu_{\st}\}$ means that the set $\bu$
is divided into two disjoint subsets. As a rule, we use roman numbers for subscripts of subsets: $\bu_{\so}$, $\bv_{\rm ii}$ etc.
However, if we deal with a big quantity of subsets, then we use standard arabic numbers for their notation. In such cases we give special comments  to avoid ambiguities.

To lighten long formulas we use a shorthand notation for products of  some functions.
Namely, if the functions $r_k$ \eqref{ratios} or the functions $g$ and $f$  depend
on sets of variables, this means that one should take the product over the corresponding set.
For example,
 \be{SH-prod}
 r_1(\bu)=\prod_{\la_k\in\bu} r_1(\la_k);\quad
 g(z, \bar w)= \prod_{w_j\in\bar w} g(z, w_j);\quad
 f(\bu,\bv)=\prod_{u_j\in\bu}\prod_{v_k\in\bv} f(u_j,v_k).
 \ee
By definition any product with respect to the empty set is equal to $1$. If we have a double product, then it is also equal to $1$ if at least
one of the sets is empty.

In section~\ref{SS-CM} we shall introduce several new scalar functions and will extend
the convention \eqref{SH-prod} to their products.

\subsection{Universal form factors\label{SS-UF}}

In this paper we reduce FFLO to the universal form factors of the monodromy matrix entries. The latter  are defined as
follows
\be{Univ-FF}
\mathfrak{F}^{(i,j)}\left(\begin{smallmatrix}
\buc & \bub \\
\bvc & \bvb
\end{smallmatrix}\right)^{a',a}_{b',b}=
\frac{ \mathbb{C}_{a',b'}(\buc;\bvc)T_{ij}(z)\mathbb{B}_{a,b}(\bub;\bvb)}
{\tau(z|\buc,\bvc)-\tau(z|\bub,\bvb)}.
\ee
Here both $\mathbb{C}_{a',b'}(\buc;\bvc)$ and $\mathbb{B}_{a,b}(\bub;\bvb)$ are on-shell
Bethe vectors, and we assume that the Bethe parameters of the two Bethe vectors are different: $\{\buc,\bvc\}\ne \{\bub,\bvb\}$.
The parameter $z$ is an arbitrary complex  number. It was proved in \cite{HutLPRS16a} that the ratio
in the r.h.s. of \eqref{Univ-FF} does not depend on $z$.

The form factors \eqref{Univ-FF} are called universal, because they are completely determined by the $R$-matrix of the model. They do not depend
on a specific representation of the $RTT$-algebra, in particular, they do not depend on the vacuum eigenvalues $\lambda_i(u)$ \eqref{vac}. In other words,
if two different integrable models are described by the $R$-matrix \eqref{DYglmn}, then they have the same universal form factors. Explicit determinant
representations for the universal form factors in the $\mathfrak{gl}(2|1)$-invariant models were obtained in \cite{HutLPRS16a}.

\section{Composite model\label{SS-CM}}

In order to access FFLO we introduce a composite model \cite{IzeK84,PakRS15b,Fuk17}.
Most naturally the composite model arises in the lattice models, where the  monodromy matrix $T(u)$  is equal to the product of local $L$-operators
\be{mat-T}
T(u)=L_M(u)\cdots L_1(u).
\ee
Here $M$ is the number of the lattice sites, and every $L$-operator satisfies $RTT$-relation with $R$-matrix \eqref{DYglmn}. Let us fix a
site $m$ ($1\le m<M$) and define two partial monodromy matrices $T^{(1)}(u)$  and $T^{(2)}(u)$ as
\be{mat-T12}
T^{(1)}(u)=L_m(u)\cdots L_1(u),\qquad T^{(2)}(u)=L_M(u)\cdots L_{m+1}(u).
\ee
Then
\be{T-TT}
T(u)=T^{(2)}(u)T^{(1)}(u).
\ee
Every $T^{(l)}(u)$ obviously satisfies $RTT$-relation \eqref{L-op-com} and has its own pseudovacuum vector $|0\rangle^{(l)}$, such that $|0\rangle= |0\rangle^{(1)}\otimes|0\rangle^{(2)}$. The operators $T_{ij}^{(2)}(u)$ and $ T_{kl}^{(1)}(v)$ supercommute with each other, as they act in different spaces.

Continuous quantum models can be obtained from the lattice ones in the limit $M\to\infty$. Obviously, the determining relation \eqref{T-TT} remains
unchanged. The partial monodromy matrices still satisfy the $RTT$ relation, and the entries of the different partial  monodromy matrices mutually supercommute.
Thus, continuous quantum models also can be considered in the framework of the composite model.

Let
\be{eigen}
T_{ii}^{(l)}(u)|0\rangle^{(l)}= \lambda_{i}^{(l)}(u)|0\rangle^{(l)}, \qquad l=1,2,
\ee
where $\lambda_{i}^{(l)}(u)$ are new free functional parameters. We also introduce
\be{rk}
r_{k}^{(l)}(u)=\frac{\lambda_{k}^{(l)}(u)}{\lambda_{2}^{(l)}(u)}, \qquad l=1,2, \qquad k=1,3.
\ee
Obviously
\be{lr}
\lambda_{i}(u)=\lambda_{i}^{(1)}(u)\lambda_{i}^{(2)}(u), \qquad
r_{k}(u)=r_{k}^{(1)}(u)r_{k}^{(2)}(u).
\ee
Below we  express form factors in terms of $r_{k}^{(1)}(u)$, therefore we introduce a special notation for
these functions
\be{ell}
r_{k}^{(1)}(u)=\ell_k(u),\qquad \text{and hence,}\qquad r_{k}^{(2)}(u)=\frac{r_k(u)}{\ell_k(u)}, \qquad k=1,3.
\ee
We extend the convention on the shorthand notation \eqref{SH-prod} to the products of the functions $\ell_k(u)$.

Any monodromy matrix in \eqref{T-TT} possesses its own Bethe vectors.
The total Bethe vector is a bilinear combination of the partial Bethe vectors \cite{Fuk17}:
\begin{equation} \label{BVcomp}
\mathbb{B}_{a,b}(\bu;\bv) = \sum r_1^{(2)}(\bu_{\so}) r_3^{(1)}(\bv_{\st}) \frac{f(\bu_{\st}, \bu_{\so})g(\bv_{\so},\bv_{\st})}{f(\bv_{\st},\bu_{\so})} \mathbb{B}_{a_2,b_2}^{(2)}(\bu_{\st}; \bv_{\st}) \mathbb{B}^{(1)}_{a_1,b_1}(\bu_{\so}; \bv_{\so}).
\end{equation}
Here the sum is taken  over all partitions $\bar u \Rightarrow \{\bu_{\so}, \bu_{\st}\}$ and $\bv \Rightarrow \{\bv_{\so}, \bv_{\st}\}$. The
cardinalities of the subsets satisfy $a_1+a_2=a$ and $b_1+b_2=b$. Recall that here we have used the convention \eqref{SH-prod} for the products
of the functions $r_k^{(\ell)}$, $f$, and $g$.

Similarly, the dual total Bethe vector is a bilinear combination of the partial dual Bethe vectors:
\begin{align} \label{dBVcomp}
&\C_{a,b}(\bu;\bv )= \sum r_1^{(1)}(\bu_{\st}) r_3^{(2)}(\bv_{\so}) \frac{f(\bu_{\so}, \bu_{\st}) g(\bv_{\st},\bv_{\so})}{f(\bv_{\so},\bu_{\st})} \C_{a_1,b_1}^{(1)}(\bu_{\so}; \bv_{\so}) \C^{(2)}_{a_2,b_2}(\bu_{\st}; \bv_{\st}).
\end{align}
Here the sum is the same as in \eqref{BVcomp}.

Observe that if the total (dual) Bethe vector is on-shell (i.e. the set $\{\bu,\bv\}$ satisfies Bethe equations), then the partial (dual) Bethe vectors
generically are not on-shell, because the subsets $\{\bu_{\so},\bv_{\so}\}$ and $\{\bu_{\st},\bv_{\st}\}$ do not satisfy Bethe equations.

Comparing these formulas with the formulas for the total (dual) Bethe vectors in $\mathfrak{gl}(3)$-based models \cite{PakRS15d} one can see that
the difference is very small. Namely, replacing the product of functions $g(\bv_{\so},\bv_{\st})$ in \eqref{BVcomp} with  the product $f(\bv_{\st},\bv_{\so})$
we obtain the expression for the total Bethe vector in the models with $\mathfrak{gl}(3)$ symmetry. Similarly, for the dual Bethe vectors
one should make the replacement $g(\bv_{\st},\bv_{\so})\;\to\;f(\bv_{\so},\bv_{\st})$.
This similarity makes it possible to calculate FFLO by the
same methods as in the case of the $\mathfrak{gl}(3)$-based models \cite{PakRS15b}.

\subsection{Zero modes\label{SS-ZM}}

We assume a standard representation of the local $L$-operators in \eqref{mat-T}:
\be{L-dep-u}
L_n(u)=\mathbf{1}+\frac cu L_n[0] +o(u^{-1}),\qquad u\to\infty.
\ee
Here $\mathbf{1}$ is the identity operator in $\mathbb{C}^{2|1}\otimes\mathcal{H}$. The matrix elements
$\bigl(L_n[0]\bigr)_{ij}$ depend on the local operators of the model. Due to \eqref{mat-T} we conclude that
the total monodromy matrix $T(u)$ and both partial monodromy matrices $T^{(l)}(u)$ have the standard expansion over $c/u$:
\be{zero-modes}
\begin{aligned}
&T(u)=\mathbf{1}+ \sum_{n=0}^\infty T[n]\,\left(\frac cu\right)^{n+1},\\
&T^{(l)}(u)=\mathbf{1}+ \sum_{n=0}^\infty T^{(l)}[n]\,\left(\frac cu\right)^{n+1},\qquad l=1,2.
\end{aligned}
\ee
The operators $T[0]$ and $T^{(l)}[0]$ respectively are called the total and the partial zero modes. Obviously, the partial zero mode $T^{(1)}[0]$ is equal to
\be{part-zm}
T^{(1)}[0]=\sum_{n=1}^m L_n[0].
\ee

Consider a form factor of the partial zero mode $T^{(1)}_{ij}[0]$
\be{FF-partZM}
\mathcal{M}^{(i,j)}\left(m\Bigr|\begin{smallmatrix}
\buc & \bub \\
\bvc & \bvb
\end{smallmatrix}\right)^{a',a}_{b',b}=\mathbb{C}_{a',b'}(\buc;\bvc)T^{(1)}_{ij}[0]\mathbb{B}_{a,b}(\bub;\bvb),
\ee
where  $\mathbb{C}_{a',b'}(\buc;\bvc)$ and $\mathbb{B}_{a,b}(\bub;\bvb)$ are on-shell
Bethe vectors. We have stressed that this form factor depends on the number  $m$ of the bulk site in \eqref{mat-T12}.
Then due to \eqref{part-zm} we obtain
\be{FF-partL}
\mathbb{C}_{a',b'}(\buc;\bvc)\bigl(L_m[0]\bigr)_{ij}\mathbb{B}_{a,b}(\bub;\bvb)=\mathcal{M}^{(i,j)}\left(m\Bigr|\begin{smallmatrix}
\buc & \bub \\ \bvc & \bvb
\end{smallmatrix}\right)^{a',a}_{b',b}- \mathcal{M}^{(i,j)}\left(m-1\Bigr|\begin{smallmatrix}
\buc & \bub \\ \bvc & \bvb
\end{smallmatrix}\right)^{a',a}_{b',b}.
\ee
Thus, knowing the form factors of the partial zero mode $T^{(1)}[0]$ we can find the form factors of the local operators
$\bigl(L_m[0]\bigr)_{ij}$. It is clear that
in the case of continuous models the finite difference in the r.h.s. of \eqref{FF-partL} turns into the derivative over a space variable.

The use of the zero modes also allows one to obtain simple relations between different form factors. It follows from the commutation relations
\eqref{TM-1} that
\be{Comm-ZM}
\begin{aligned}
 {}[T_{ij}[0],T_{kl}[0]\} & = (-1)^{[i][j]+[i][l]+[j][l]}(\delta_{il} T_{kj}[0] -\delta_{kj} T_{il}[0]),\\
[T_{ij}^{(s)}[0],T_{kl}^{(s)}[0]\} & = (-1)^{[i][j]+[i][l]+[j][l]}(\delta_{il} T_{kj}^{(s)}[0] -\delta_{kj} T_{il}^{(s)}[0]),
\qquad s=1,2.\\
\end{aligned}
\ee
Using $T_{ij}[0]=T_{ij}^{(1)}[0]+T_{ij}^{(2)}[0]$ and $[T_{ij}^{(1)}[0],T_{kl}^{(2)}[0]\} =0$ we arrive at
\be{Comm-ZM12}
[T_{ij}^{(1)}[0],T_{kl}[0]\}  = (-1)^{[i][j]+[i][l]+[j][l]}(\delta_{il} T_{kj}^{(1)}[0] -\delta_{kj} T_{il}^{(1)}[0]).
\ee
Equation \eqref{Comm-ZM12} yields
\begin{multline}\label{FFlk-FFij}
\delta_{il} \mathcal{M}^{(k,j)}\left(m\Bigr|\begin{smallmatrix}
\buc & \bub \\ \bvc & \bvb
\end{smallmatrix}\right)^{a',a}_{b',b}- \delta_{kj} \mathcal{M}^{(i,l)}\left(m\Bigr|\begin{smallmatrix}
\buc & \bub \\ \bvc & \bvb
\end{smallmatrix}\right)^{a',a}_{b',b}\\
=(-1)^{[i][j]+[i][l]+[j][l]}\mathbb{C}_{a',b'}(\buc;\bvc)[T_{ij}^{(1)}[0],T_{kl}[0]\}\mathbb{B}_{a,b}(\bub;\bvb).
\end{multline}
The actions of the total zero modes $T_{kl}[0]$ onto (dual) on-shell Bethe vectors were studied in \cite{HutLPRS16a}. Under
this action a (dual) on-shell vector either vanishes or remains on-shell. Therefore, the expectation value  in the r.h.s.
of \eqref{FFlk-FFij} is related to the form factor of the partial zero mode $T_{ij}^{(1)}[0]$. Equation \eqref{FFlk-FFij}, thus,
allows us to express this form factor in terms of $\mathcal{M}^{(k,j)}$ and $\mathcal{M}^{(i,l)}$. We consider specific examples
of these relationships in section~\ref{S-FFOD}.

\section{Main results\label{S-MR}}

We have shown in the previous section that FFLO can be reduced to the form factors of the partial zero
modes $T^{(1)}_{ij}[0]$. Studying these form factors one should distinguish between two cases. In the first case an on-shell Bethe vector
$\B_{a,b}(\bub;\bvb)$ and a dual on shell Bethe vector $\C_{a',b'}(\buc;\bvc)$ correspond to different eigenvalues of the transfer
matrix. Then we say that $\{\buc,\bvc\} \neq \{\bub,\bvb\}$. Overwise, if $\{\buc,\bvc\} =\{\bub,\bvb\}$, then both vectors
correspond to the same eigenvalue. The latter case occurs for form factors of the diagonal elements $T^{(1)}_{ii}[0]$ only.
\begin{thm}
Let $\B_{a,b}(\bub;\bvb)$ and $\C_{a',b'}(\buc;\bvc)$ be on-shell (dual) Bethe vectors such that $\{\buc,\bvc\} \neq \{\bub,\bvb\}$. Then
\begin{equation}\label{FF-gener}
  \mathcal{M}^{(i,j)}
\left( m\Bigr|\begin{smallmatrix}
  \buc & \bub \\ \bvc & \bvb
\end{smallmatrix}\right)^{a',a}_{b',b}
%
  = \left( \frac{\ell_1(\buc)\ell_3(\bvb)}{\ell_1(\bub)\ell_3(\bvc)} -1 \right) \mathfrak{F}^{(i,j)}
\left( \begin{smallmatrix}
  \buc & \bub \\ \bvc & \bvb
\end{smallmatrix}\right)^{a',a}_{b',b},
\end{equation}
where $\mathfrak{F}^{(i,j)}$ is the universal form factor of the total monodromy matrix element $T_{ij}(z)$.
\end{thm}

\begin{thm} \label{FF-diag}
Let $\B_{a,b}(\bu;\bv)$ be an on-shell Bethe vector and $\C_{a,b}(\bu;\bv)$ be its dual on-shell Bethe vector. Let $\C_{a,b}(\bu(\bar\kappa);\bv(\bar\kappa))$ be a twisted on-shell deformation of $\C_{a,b}(\bu;\bv)$ such that $\bu(\bar\kappa)=\bu$, $\bv(\bar\kappa)=\bv$ at\footnote{Here and below
$\bar\kappa=\{\kappa_1,\kappa_2,\kappa_3\}$, and $\bar\kappa=1$ means $\kappa_1=\kappa_2=\kappa_3=1$.} $\bar\kappa=1$. Then
\begin{equation}\label{FFdiag-eq}
\mathcal{M}^{(i,i)}
\left( m\Bigr|\begin{smallmatrix}
  \bu & \bu \\ \bv & \bv
\end{smallmatrix}\right)^{a,a}_{b,b}
=  \left(  \lambda^{(1)}_i[0]
 + (-1)^{[i]}  \frac{\mathrm{d}}{\mathrm{d} \kappa_i} \log \frac{\ell_1(\bu(\bar\kappa))}{\ell_3(\bv(\bar\kappa))} \Big|_{\bar \kappa=1}   \right) ||\B_{a,b}(\bu;\bv)||^2,
\end{equation}
where $\lambda^{(1)}_k[0]$ can be found from the expansion
\be{zero-modesl}
\lambda^{(1)}_k(u)=1+ \lambda^{(1)}_k[0]\,\frac cu +\dots\;,\qquad u\to\infty.
\ee
\end{thm}

We prove theorems~\ref{FF-gener} and~\ref{FF-diag} in section~\ref{S-FFOD}. The most technical part of the proof is
given in appendix~\ref{A-CALC}.  Here we would like to mention only that
the general strategy of the proof is the same as in the $\mathfrak{gl}(3)$ case \cite{PakRS15b}.

Comparing these formulas with the corresponding expressions in the $\mathfrak{gl}(3)$-based models \cite{PakRS15b} we see that the only
difference is the sign factor $(-1)^{[i]}$ in \eqref{FFdiag-eq}. Certainly, the determinant formulas for the universal form factors in the
models with $\mathfrak{gl}(2|1)$ and $\mathfrak{gl}(3)$ symmetries are different, however, the relation between $\mathcal{M}^{(i,j)}$
and $\mathfrak{F}^{(i,j)}$ is the same (modulus the sign factor mentioned above). Most probably, the same relation takes place in the
general $\mathfrak{gl}(m|n)$ case as well. One should  remember, however, that in models with $\mathfrak{gl}(2|1)$-invariant $R$-matrix there exist compact determinant representations for the universal form factors \cite{HutLPRS16a}. These representations can be directly used for analysis of correlation functions.
At the same time, analogous determinant formulas for the universal form factors
in the general $\mathfrak{gl}(m|n)$ case are unknown up to date.

Observe that the dependence on the local site $m$ in \eqref{FF-gener}, \eqref{FFdiag-eq} enters only the functions $\ell_k(u)$ and $\lambda^{(1)}_k[0]$. This dependence follows
from representation \eqref{mat-T12}
\begin{equation}
\lambda^{(1)}_k(u) = \prod_{n=1}^m \lambda_k(u|n), \qquad k=1,2,3,
\end{equation}
\be{lkm}
\ell_k(u)=\prod_{n=1}^m \frac{\lambda_k(u|n)}{\lambda_2(u|n)},\qquad k=1,3,
\ee
where  $\lambda_k(u|n)$ are vacuum eigenvalues of the local $L$-operators entries $\bigl(L_n(u)\bigr)_{kk}$.
Due to  \eqref{L-dep-u} the expansion of the vacuum eigenvalues $\lambda_k(u|n)$ takes the form
\be{exp-ve}
\lambda_k(u|n) = 1+  \lambda_k[0|n] \frac{c}{u} + \dots ,\qquad u\to\infty,
\ee
and hence, the coefficient  $\lambda^{(1)}_k[0]$ in \eqref{zero-modesl} is
\begin{equation}
\lambda^{(1)}_k[0] = \sum_{n=1}^m \lambda_k[0|n] .
\end{equation}
Then, due to \eqref{FF-partL} we find for $\{\buc,\bvc\} \neq \{\bub,\bvb\}$
\be{FF-partL-answ}
\mathbb{C}_{a',b'}(\buc;\bvc)\bigl(L_m[0]\bigr)_{ij}\mathbb{B}_{a,b}(\bub;\bvb)=\bigl(\mathcal{L}_m-1\bigr)
\Bigl(\prod_{n=1}^{m-1}\mathcal{L}_n\Bigr)\;
\mathfrak{F}^{(i,j)}
\left( \begin{smallmatrix}
  \buc & \bub \\ \bvc & \bvb
\end{smallmatrix}\right)^{a',a}_{b',b},
\ee
where
\be{cLn}
\mathcal{L}_n=\frac{\ell_1(\buc|n)\ell_3(\bvb|n)}{\ell_1(\bub|n)\ell_3(\bvc|n)}.
\ee
If $\{\buc,\bvc\} =\{\bub,\bvb\}=\{\bu,\bv\}$, then
\begin{equation}\label{FFdiag-answ}
\C_{a,b}(\bu;\bv) \bigl(L_m[0]\bigr)_{ii} \B_{a,b}(\bu;\bv)
= \Bigl[\lambda_i[0|m]
+  (-1)^{[i]}\frac{\mathrm{d}}{\mathrm{d} \kappa_i} \log \frac{\ell_1\bigl(\bu(\bar\kappa)|m\bigr)}
{\ell_3\bigl(\bv(\bar\kappa)|m\bigr)} \Big|_{\bar \kappa=1} \Bigr]  ||\B_{a,b}(\bu;\bv)||^2,
\end{equation}
where $\bu(\bar\kappa)$ and $\bv(\bar\kappa)$ are the deformations of $\bu$ and $\bv$ described in theorem~\ref{FF-diag}.

\section{Generating functional for form factors of  partial zero modes\label{S-FFOD}}

All the form factors of the partial zero modes $T^{(1)}_{ij}[0]$ can be found from a special generating functional.
Consider an operator
\be{op-Q}
Q_{\bar\beta}=\sum_{i=1}^3 (-1)^{[i]}  \beta_i T^{(1)}_{ii}[0],
\ee
where $\beta_i$ are some complex numbers.

Let  $\mathbb{B}_{a,b}(\bub;\bvb)$ be an on-shell Bethe vector.
Let also $\mathbb{C}^{(\kappa)}_{a,b}(\buc;\bvc)$ be a twisted dual on-shell Bethe vector with the twist  $\kappa=\diag\{\kappa_1,\kappa_2,\kappa_3\}$.
We stressed this fact by adding the superscript $(\kappa)$ to the vector $\mathbb{C}_{a,b}(\buc;\bvc)$. Suppose that
$\kappa_i=e^{\beta_i}$ and consider the following expectation value
\be{genfun-def}
\mathcal{M}^{(\kappa)}\left(m\Bigr|\begin{smallmatrix}
\buc & \bub \\
\bvc & \bvb
\end{smallmatrix}\right)^{a}_{b}
= \mathbb{C}^{(\kappa)}_{a,b}(\buc;\bvc)\; e^{Q_{\bar\beta}}\;\mathbb{B}_{a,b}(\bub;\bvb).
\ee
Taking the derivative of this generating functional over $\beta_i$ at $\bar\kappa=1$ (that is, all $\beta_j=0$) we obtain
\be{genfun-der}
(-1)^{[i]}\mathcal{M}^{(i,i)}\left(m\Bigr|\begin{smallmatrix}
\buc & \bub \\
\bvc & \bvb
\end{smallmatrix}\right)^{a,a}_{b,b}=\frac{\mathrm{d}}{\mathrm{d} \beta_i}\Bigl[
\mathcal{M}^{(\kappa)}\left(m\Bigr|\begin{smallmatrix}
\buc & \bub \\
\bvc & \bvb
\end{smallmatrix}\right)^{a}_{b}
- \mathbb{C}^{(\kappa)}_{a,b}(\buc;\bvc)\mathbb{B}_{a,b}(\bub;\bvb)\Bigr]_{\bar\kappa=1}.
\ee
It was shown in \cite{HutLPRS16b}  that for $\{\buc,\bvc\}\bigr|_{\bar\kappa=1} \neq \{\bub,\bvb\}$
\be{der-SP}
\frac{\mathrm{d}}{\mathrm{d} \beta_i}
 \mathbb{C}^{(\kappa)}_{a,b}(\buc;\bvc)\mathbb{B}_{a,b}(\bub;\bvb)\Bigr|_{\bar\kappa=1}=
(-1)^{[i]}\mathfrak{F}^{(i,i)}
\left( \begin{smallmatrix}
  \buc & \bub \\ \bvc & \bvb
\end{smallmatrix}\right)^{a,a}_{b,b}.
\ee
Hence, we obtain in this case
\be{genfun-FF1}
\mathcal{M}^{(i,i)}\left(m\Bigr|\begin{smallmatrix}
\buc & \bub \\
\bvc & \bvb
\end{smallmatrix}\right)^{a,a}_{b,b}=(-1)^{[i]}\frac{\mathrm{d}}{\mathrm{d} \beta_i}
\mathcal{M}^{(\kappa)}\left(m\Bigr|\begin{smallmatrix}
\buc & \bub \\
\bvc & \bvb
\end{smallmatrix}\right)^{a}_{b}\Bigr|_{\bar\kappa=1}
-\mathfrak{F}^{(i,i)}
\left( \begin{smallmatrix}
  \buc & \bub \\ \bvc & \bvb
\end{smallmatrix}\right)^{a,a}_{b,b}.
\ee
Thus, calculating the generating functional
\eqref{genfun-def}, we can find the form factors of the diagonal partial zero modes $T^{(1)}_{ii}[0]$ at least
for $\{\buc,\bvc\}\bigr|_{\bar\kappa=1} \neq \{\bub,\bvb\}$. The case $\{\buc,\bvc\}\bigr|_{\bar\kappa=1} =\{\bub,\bvb\}$ will be considered later.

The form factors of the partial zero modes $T^{(1)}_{ij}[0]$ with $i\ne j$ can be obtained via relation \eqref{FFlk-FFij}. Let us give an example.
Let $i=j=l=2$ and $k=1$ in \eqref{FFlk-FFij}. Then this formula takes the form
\begin{equation}\label{FF12-FF22}
\mathcal{M}^{(1,2)}\left(m\Bigr|\begin{smallmatrix}
\buc & \bub \\ \bvc & \bvb
\end{smallmatrix}\right)^{a+1,a}_{b,b}
=\mathbb{C}_{a+1,b}(\buc;\bvc)\bigl(T_{22}^{(1)}[0]T_{12}[0]-T_{12}[0]T_{22}^{(1)}[0]\bigr)\mathbb{B}_{a,b}(\bub;\bvb).
\end{equation}
Due to the results of \cite{HutLPRS16a} we have
\be{act-ZM}
\mathbb{C}_{a+1,b}(\buc;\bvc)T_{12}[0]=0, \qquad T_{12}[0]\mathbb{B}_{a,b}(\bub;\bvb)=\lim_{w\to\infty}
\frac wc\mathbb{B}_{a+1,b}(\{w,\bub\};\bvb),
\ee
where we used the fact that both $\mathbb{C}_{a+1,b}(\buc;\bvc)$ and $\mathbb{B}_{a,b}(\bub;\bvb)$ are on-shell. Due to the Bethe equations
\eqref{BE} we conclude that if $\mathbb{B}_{a,b}(\bub;\bvb)$ is on-shell, then $\mathbb{B}_{a+1,b}(\{w,\bub\};\bvb)$ is also on-shell
at $w\to\infty$. This is because $r_1(w)\to 1$ at $w\to\infty$ according to  expansion \eqref{zero-modes}. Thus, we arrive at
\begin{equation}\label{FF12-FF22-1}
\mathcal{M}^{(1,2)}\left(m\Bigr|\begin{smallmatrix}
\buc & \bub \\ \bvc & \bvb
\end{smallmatrix}\right)^{a+1,a}_{b,b}
=\lim_{w\to\infty}
\frac wc\mathbb{C}_{a+1,b}(\buc;\bvc)T_{22}^{(1)}[0]\mathbb{B}_{a+1,b}(\{w,\bub\};\bvb),
\end{equation}
and since both vectors in the r.h.s. of \eqref{FF12-FF22-1} are on-shell, we obtain
\begin{equation}\label{FF12-FF22-2}
\mathcal{M}^{(1,2)}\left(m\Bigr|\begin{smallmatrix}
\buc & \bub \\ \bvc & \bvb
\end{smallmatrix}\right)^{a+1,a}_{b,b}
=\lim_{w\to\infty}
\frac wc\mathcal{M}^{(2,2)}\left(m\Bigr|\begin{smallmatrix}
\buc & \{w,\bub\} \\ \bvc & \bvb
\end{smallmatrix}\right)^{a+1,a+1}_{b,b}.
\end{equation}
Thus, knowing an explicit representation for the form factor $\mathcal{M}^{(2,2)}$ we can find the form factor
$\mathcal{M}^{(1,2)}$ sending one of the Bethe parameters to infinity. Similarly all the other form factors of the off-diagonal
partial zero modes $T_{ij}^{(1)}[0]$ can be found.

\subsection{Calculation of the generating functional\label{SS-CGF}}

The calculation of the generating functional \eqref{genfun-def} is straightforward. First of all, we use explicit expressions
\eqref{BVcomp} and \eqref{dBVcomp} for the total (dual) Bethe vectors in terms of the partial ones. This allows us to find the action
of the operator $\exp\{Q_{\bar\beta}\}$ onto the Bethe vector $\mathbb{B}_{a,b}(\bub;\bvb)$. After this we obtain a sum over partitions
of the Bethe parameters involving two scalar products of the partial Bethe vectors. Using an explicit representation for the scalar
product of generic Bethe vectors in the $\mathfrak{gl}(2|1)$-based models \cite{HutLPRS16c} we find an explicit expression for the
generating functional in terms of a sum over partitions. This sum can be further simplified leading eventually to the final result.
We describe here the first steps of this derivation. The most technical part is shifted to appendix~\ref{A-CALC}.

We start with equation \eqref{BVcomp} for the total Bethe vector.
If this vector is on-shell, then we can present the product of functions $r_{1}^{(2)}(\bub_{\so})$ as
$r_{1}^{(2)}(\bub_{\so})=r_1(\bub_{\so})\ell^{-1}_1(\bub_{\so})$ (see \eqref{ell}) and express $r_1(\bub_{\so})$ in terms of Bethe equations. Then we obtain
\be{BV-BV-1}
\mathbb{B}_{a,b}(\bub;\bvb)=\sum \frac{\ell_{3}(\bvb_{\st})}{\ell_{1}(\bub_{\so})} f(\bub_{\so},\bub_{\st})
g(\bvb_{\so},\bvb_{\st})f(\bvb_{\so},\bub_{\so})\;
\mathbb{B}_{a_{\st},b_{\st}}^{(2)}(\bub_{\st};\bvb_{\st})\mathbb{B}_{a_{\so},b_{\so}}^{(1)}(\bub_{\so};\bvb_{\so}) .
\ee
Similarly, if $\mathbb{C}^{(\kappa)}_{a,b}(\buc;\bvc)$ is a twisted on-shell Bethe vector, then we can present the product of functions $r_{3}^{(2)}(\bvc_{\so})$ as
$r_{3}^{(2)}(\bvc_{\so})=r_3(\bvc_{\so})\ell^{-1}_3(\bvc_{\so})$  and express $r_3(\bvc_{\so})$ in terms of the twisted Bethe equations. Then we have
\be{CV-CV-eig}
\mathbb{C}^{(\kappa)}_{a,b}(\buc;\bvc)=\sum \left(\frac{\kappa_2}{\kappa_3}\right)^{b_{\so}}\frac{\ell_{1}(\buc_{\st})} {\ell_{3}(\bvc_{\so})}f(\buc_{\so},\buc_{\st}) g(\bvc_{\st},\bvc_{\so})f(\bvc_{\so},\buc_{\so})\;
\mathbb{C}_{a_{\so},b_{\so}}^{(1)}(\buc_{\so};\bvc_{\so}) \mathbb{C}_{a_{\st},b_{\st}}^{(2)}(\buc_{\st};\bvc_{\st}).
\ee

Now we should compute the action of $\exp\{Q_{\bar\beta}\}$ onto the Bethe vector $\mathbb{B}_{a,b}(\bub;\bvb)$. Obviously,  the partial zero modes
$T^{(1)}_{ii}[0]$ act only on the partial Bethe vectors $\mathbb{B}^{(1)}_{a_{\so},b_{\so}}$ and do not act on $\mathbb{B}^{(2)}_{a_{\st},b_{\st}}$.
The explicit action formulas are \cite{HutLPRS16a}:
\be{actBV}
\begin{aligned}
T^{(1)}_{11}[0]\mathbb{B}^{(1)}_{a_{\so},b_{\so}}(\bub_{\so};\bvb_{\so})&=(\lambda^{(1)}_1[0]-a_{\so})\mathbb{B}^{(1)}_{a_{\so},b_{\so}}(\bub_{\so};\bvb_{\so}), \\
T^{(1)}_{22}[0]\mathbb{B}^{(1)}_{a_{\so},b_{\so}}(\bub_{\so};\bvb_{\so})&=(\lambda_2^{(1)}[0]+a_{\so}-b_{\so})\mathbb{B}^{(1)}_{a_{\so},b_{\so}}(\bub_{\so};\bvb_{\so}), \\
T^{(1)}_{33}[0]\mathbb{B}^{(1)}_{a_{\so},b_{\so}}(\bub_{\so};\bvb_{\so})&=(\lambda^{(1)}_3[0] -b_{\so})\;\mathbb{B}^{(1)}_{a_{\so},b_{\so}}(\bub_{\so};\bvb_{\so}).
\end{aligned}
\ee
Here $\mathbb{B}^{(1)}_{a_{\so},b_{\so}}(\bub_{\so};\bvb_{\so})$ is a generic partial Bethe vector. Thus, we find
\begin{multline}\label{act-BV-BV}
e^{Q_{\bar\beta}}\mathbb{B}_{a,b}(\bub;\bvb)=\sum  e^{\mathcal{Q}_{\bar\beta}+ a_{\so}(\beta_2-\beta_1) +b_{\so}(\beta_3-\beta_2)}
\frac{\ell_{3}(\bvb_{\st})}{\ell_{1}(\bub_{\so})} \\
\times f(\bub_{\so},\bub_{\st}) g(\bvb_{\so},\bvb_{\st}) f(\bvb_{\so},\bub_{\so})
\mathbb{B}_{a_{\st},b_{\st}}^{(2)}(\bub_{\st};\bvb_{\st})\mathbb{B}_{a_{\so},b_{\so}}^{(1)}(\bub_{\so};\bvb_{\so}) ,
\end{multline}
where
\be{qb}
\mathcal{Q}_{\bar\beta}=\sum_{i=1}^3(-1)^{[i]}\beta_i\lambda_i^{(1)}[0]
\ee
is the eigenvalue of the operator $Q_{\bar\beta}$ on the vector $|0\rangle$.
Substituting \eqref{CV-CV-eig} and \eqref{act-BV-BV} into \eqref{genfun-def} we arrive at
\begin{multline}\label{Mat-elS}
\mathcal{M}^{(\kappa)}\left(m\Bigr|\begin{smallmatrix}
\buc & \bub \\
\bvc & \bvb
\end{smallmatrix}\right)^{a}_{b}=\sum
e^{\mathcal{Q}_{\bar\beta}+ a_{\so}(\beta_2-\beta_1)}
\frac{\ell_1(\buc_{\st})\ell_3(\bvb_{\st})}{\ell_1(\bub_{\so})\ell_3(\bvc_{\so})}\;
f(\buc_{\so},\buc_{\st})f(\bub_{\so},\bub_{\st})g(\bvc_{\st},\bvc_{\so}) g(\bvb_{\so},\bvb_{\st})\\
\times
f(\bvb_{\so},\bub_{\so})f(\bvc_{\so},\buc_{\so})
\; \mathbb{C}_{a_{\so},b_{\so}}^{(1)}(\buc_{\so};\bvc_{\so})\mathbb{B}_{a_{\so},b_{\so}}^{(1)}(\bub_{\so};\bvb_{\so}) \cdot \mathbb{C}_{a_{\st},b_{\st}}^{(2)}(\buc_{\st};\bvc_{\st})\mathbb{B}_{a_{\st},b_{\st}}^{(2)}(\bub_{\st};\bvb_{\st}).
\end{multline}
Thus, the problem of calculating the generating functional is reduced to the calculation of the scalar products and further summation over partitions.
Further derivation is quite technical, therefore, we give the details in appendix~\ref{A-CALC}. We would like to mention only that this derivation
goes along the same lines as in the $\mathfrak{gl}(3)$-case \cite{PakRS15b} with minor modifications. Here we formulate the final result only.

\begin{prop} \label{thmM}
Let $\mathbb{B}_{a,b}(\bub;\bvb)$ be an on-shell Bethe vector and $\mathbb{C}^{(\kappa)}_{a,b}(\buc;\bvc)$ be a twisted dual on-shell Bethe vector.
Then
\be{result}
\mathcal{M}^{(\kappa)}\left(m\Bigr|\begin{smallmatrix}
\buc & \bub \\
\bvc & \bvb
\end{smallmatrix}\right)^{a}_{b}= e^{\mathcal{Q}_{\bar\beta}}
\frac{\ell_1(\buc)\ell_3(\bvb)}{\ell_1(\bub)\ell_3(\bvc)}\;\mathbb{C}^{(\kappa)}_{a,b}(\buc;\bvc)\mathbb{B}_{a,b}(\bub;\bvb).
\ee
\end{prop}

Suppose that $\{\buc,\bvc\}\bigr|_{\bar\kappa=1} \neq \{\bub,\bvb\}$. Then, it follows immediately from \eqref{result} that
\be{der-result}
\frac{\mathrm{d}}{\mathrm{d} \beta_i}\mathcal{M}^{(\kappa)}\left(m\Bigr|\begin{smallmatrix}
\buc & \bub \\
\bvc & \bvb
\end{smallmatrix}\right)^{a}_{b}\Bigr|_{\bar\kappa=1}=
\frac{\ell_1(\buc)\ell_3(\bvb)}{\ell_1(\bub)\ell_3(\bvc)}\;
\frac{\mathrm{d}}{\mathrm{d} \beta_i}\mathbb{C}^{(\kappa)}_{a,b}(\buc;\bvc)\mathbb{B}_{a,b}(\bub;\bvb)\Bigr|_{\bar\kappa=1}.
\ee
Indeed, for $\bar\kappa=1$ the scalar product in \eqref{result} turns into the scalar product of two different on-shell Bethe vectors. Hence,
it vanishes. Therefore, the $\beta_i$-derivative must act on this scalar product only, otherwise we obtain zero contribution. Using then
\eqref{der-SP} and \eqref{genfun-FF1} we immediately obtain
\begin{equation}\label{FFii-gener}
  \mathcal{M}^{(i,i)}
\left( m\Bigr|\begin{smallmatrix}
  \buc & \bub \\ \bvc & \bvb
\end{smallmatrix}\right)^{a,a}_{b,b}
%
  = \left( \frac{\ell_1(\buc)\ell_3(\bvb)}{\ell_1(\bub)\ell_3(\bvc)} -1 \right) \mathfrak{F}^{(i,i)}
\left( \begin{smallmatrix}
  \buc & \bub \\ \bvc & \bvb
\end{smallmatrix}\right)^{a,a}_{b,b},
\end{equation}
for $\{\buc,\bvc\} \neq \{\bub,\bvb\}$. Thus, we reproduce \eqref{FF-gener} for $i=j$. The form factors of the off-diagonal partial zero modes
then can be derived via \eqref{FFlk-FFij}.

Consider finally the case $\{\buc,\bvc\}\bigr|_{\bar\kappa=1} = \{\bub,\bvb\}=\{\bu,\bv\}$. Then \eqref{result} yields
\begin{multline}\label{genfun-FF10}
\frac{\mathrm{d}}{\mathrm{d} \beta_i}
\mathcal{M}^{(\kappa)}\left(m\Bigr|\begin{smallmatrix}
\bu & \bu \\
\bv & \bv
\end{smallmatrix}\right)^{a}_{b}\Bigr|_{\bar\kappa=1}= \frac{\mathrm{d}}{\mathrm{d} \beta_i}\mathbb{C}^{(\kappa)}_{a,b}(\bu(\bar\kappa);\bv(\bar\kappa))\mathbb{B}_{a,b}(\bu;\bv)\Bigr|_{\bar\kappa=1}
\\
+
\left( (-1)^{[i]}\lambda_i^{(1)}[0]  +  \frac{\mathrm{d}}{\mathrm{d} \beta_i} \log \frac{\ell_1(\bu(\bar\kappa))}{\ell_3(\bv(\bar\kappa))} \Big|_{\bar \kappa=1}   \right) ||\B_{a,b}(\bu;\bv)||^2.
\end{multline}
Comparing this equation with \eqref{genfun-der} we arrive at
\begin{equation}\label{FFdiag-eq0}
\mathcal{M}^{(i,i)}
\left( m\Bigr|\begin{smallmatrix}
  \bu & \bu \\ \bv & \bv
\end{smallmatrix}\right)^{a,a}_{b,b}
=  \left( \lambda_i^{(1)}[0]
  + (-1)^{[i]}  \frac{\mathrm{d}}{\mathrm{d} \beta_i} \log \frac{\ell_1(\bu(\bar\kappa))}{\ell_3(\bv(\bar\kappa))} \Big|_{\bar \kappa=1}   \right) ||\B_{a,b}(\bu;\bv)||^2.
\end{equation}
Finally, using
\be{der-der}
\frac{\mathrm{d}}{\mathrm{d} \beta_i}\Big|_{\bar \kappa=1}=\frac{\mathrm{d}}{\mathrm{d} \log\kappa_i}\Big|_{\bar \kappa=1}
=\frac{\mathrm{d}}{\mathrm{d} \kappa_i}\Big|_{\bar \kappa=1},
\ee
we reproduce \eqref{FFdiag-eq}.

\section*{Conclusion}

In this paper we have calculated FFLO in the integrable models with $\mathfrak{gl}(2|1)$-invariant
$R$-matrix. We have shown that these form factors are proportional to the universal form factors. Our derivation is not based
on a specific representation of the $RTT$-algebra, and thus, it is valid for a wide class of integrable models solvable by the
algebraic Bethe ansatz (besides the t-J model see e.g. \cite{BraGLZ95,BedF95,RamM96,PfaF96}). Due to an isomorphism between
$Y\bigl(\mathfrak{gl}(2|1)\bigr)$ and $Y\bigl(\mathfrak{gl}(1|2)\bigr)$ these results also can be applied to the
$\mathfrak{gl}(1|2)$-invariant models.

It is worth mentioning that for calculating the FFLO it was enough to find a special generating functional
\eqref{genfun-def}. All the form factors then can be found by taking $\kappa$-derivatives of this generating functional and sending
some of the Bethe parameters to infinity. These close relations between different form factors occur due to the commutation relations between
zero modes $T_{ij}[0]$ in the expansion \eqref{zero-modes}.
Generically, local $L$-operators not necessarily have the form \eqref{L-dep-u}, therefore, the monodromy matrix
might have an asymptotic expansion different from \eqref{zero-modes}.
This may lead to a redefinition of the zero modes and changing of the commutation relations between them, which in turn
yields a modification of the relations between the FFLO (see e.g. \cite{PakRS15c}). Nevertheless,
the latter still are proportional to the universal form factors. Therefore, the results of this paper for the most part apply
to the models in which the monodromy matrix does not have the asymptotic expansion \eqref{zero-modes}.

Our results are in complete analogue with the ones obtained in \cite{PakRS15b} for the models with the
$\mathfrak{gl}(3)$-invariant $R$-matrix. One can conjecture that the relationship between the FFLO
and the universal form factors remains true in the general $\mathfrak{gl}(m|n)$ case. However, compact representations for the universal
form factors in the $\mathfrak{gl}(m|n)$-invariant models are not known for today. At the same time, in the models with
$\mathfrak{gl}(2|1)$ or $\mathfrak{gl}(1|2)$ symmetries the universal form factors were calculated in \cite{HutLPRS16a} in terms
of determinants. These representations allow one to use our results for studying correlation functions via the form factor expansion.

\section*{Acknowledgements}

The work of J.F. has been supported by the Grant Agency of the Czech Technical
University in Prague, grant No. $SGS15/215/OHK4/3T/14$, and by the Grant of
the Plenipotentiary of the Czech Republic at JINR, Dubna. He is also grateful to A. P. Isaev and \v{C}. Burd\'{i}k for discussions and
support.

A part of this work, section~\ref{S-FFOD}, was performed in Steklov Mathematical Institute
of Russian Academy of Sciences by N.A.S. and he was supported by the Russian Science Foundation
under grant 14-50-00005.
\appendix

\section{Summation over partitions\label{A-CALC}}

\subsection{Scalar products}

In this section we give some results of the papers \cite{HutLPRS16c,HutLPRS16b}.
The scalar product of generic Bethe vectors in $\mathfrak{gl}(2|1)$-based models is given by the formula
 \begin{multline}\label{Resh-SP}
 \mathbb{C}_{a,b}(\buc;\bvc)\mathbb{B}_{a,b}(\bub;\bvb)=\sum r_1(\bub_{\so})r_1(\buc_{\st})
 r_3(\bvc_{\st})r_3(\bvb_{\so})
  f(\buc_{\so},\buc_{\st})  f(\bub_{\st},\bub_{\so})     \num
 \times  g(\bvc_{\st},\bvc_{\so})   g(\bvb_{\so},\bvb_{\st})\frac{f(\bvc_{\so},\buc_{\so})f(\bvb_{\st},\bub_{\st})}{f(\bvc,\buc)f(\bvb,\bub)}
  \;Z_{a_{\st},b_{\so}}(\buc_{\st};\bub_{\st}|\bvc_{\so};\bvb_{\so})
 Z_{a_{\so},b_{\st}}(\bub_{\so};\buc_{\so}|\bvb_{\st};\bvc_{\st}).
 \end{multline}
Here the sum is taken over the partitions
 \be{part-1}
 \begin{array}{ll}
 \buc\Rightarrow\{\buc_{\so},\;\buc_{\st}\}, &\qquad  \bvc\Rightarrow\{\bvc_{\so},\;\bvc_{\st}\},\\
 \bub\Rightarrow\{\bub_{\so},\;\bub_{\st}\}, &\qquad  \bvb\Rightarrow\{\bvb_{\so},\;\bvb_{\st}\} .
 \end{array}
 \ee
The partitions are independent except that $\#\bub_{\so}=\#\buc_{\so}=a_{\so}$ with $a_{\so}=0,\dots,a$, 
and $\#\bvb_{\so}=\#\bvc_{\so}=b_{\so}$ with $b_{\so}=0,\dots,b$.

The rational functions $Z_{a_{\st},b_{\so}}(\buc_{\st},\bub_{\st}|\bvc_{\so},\bvb_{\so})$ and $Z_{a_{\so},b_{\st}}(\bub_{\so},\buc_{\so}|\bvb_{\st},\bvc_{\st})$ are so called highest coefficients. Explicit determinant formulas for them can be found in \cite{HutLPRS16c}. We do not use these explicit presentations in our calculations.

Equation \eqref{Resh-SP} holds for arbitrary Bethe vectors. In other words, we do not assume any constraint between functional parameters $r_k$ and complex variables $\bucb$ and $\bvcb$. However, one can consider particular cases of \eqref{Resh-SP}, where certain constraints are imposed. One of these particular cases is the scalar product of the twisted on-shell and usual on-shell Bethe vectors. Then the functional parameters $r_k$ can be expressed in terms
$\bucb$ and $\bvcb$ via (twisted) Bethe equations \eqref{BE}, \eqref{TBE}.
Denoting this scalar product by  $\mathcal{S}_{a,b}^{(\kappa)}$ we obtain
 \begin{multline}\label{Resh-SP-eig}
 \mathcal{S}_{a,b}^{(\kappa)}=\sum  \left(\frac{\kappa_2}{\kappa_1}\right)^{a_{\st}}
 \left(\frac{\kappa_2}{\kappa_3}\right)^{b_{\st}}
  f(\buc_{\st},\buc_{\so})  f(\bub_{\so},\bub_{\st})  g(\bvc_{\so},\bvc_{\st})
   g(\bvb_{\st},\bvb_{\so})
  \num
 \times f(\bvc_{\st},\buc_{\st})f(\bvb_{\so},\bub_{\so}) \;Z_{a_{\st},b_{\so}}(\buc_{\st};\bub_{\st}|\bvc_{\so};\bvb_{\so})
 Z_{a_{\so},b_{\st}}(\bub_{\so};\buc_{\so}|\bvb_{\st};\bvc_{\st}).
 \end{multline}
 {\sl Remark 1.} We would like to draw attention of the reader that here the parameters $\bucb$ and $\bvcb$ still are arbitrary complex numbers,
in spite of we used (twisted) Bethe equations to obtain \eqref{Resh-SP-eig}. The matter is that the functions $r_k$ are free functional parameters.
Therefore, the (twisted) Bethe equations give the constraints for these functional parameters, but not for the Bethe parameters $\bucb$ and $\bvcb$.

Setting $\bar\kappa=1$ we obtain the scalar product of two on-shell Bethe vectors, which vanishes for $a+b>0$. Hence,
 \begin{multline}\label{Resh-SP-eig1}
\delta_{a+b,0}=\sum    f(\buc_{\st},\buc_{\so})  f(\bub_{\so},\bub_{\st})   g(\bvc_{\so},\bvc_{\st})
   g(\bvb_{\st},\bvb_{\so})
  \num
 \times f(\bvc_{\st},\buc_{\st})f(\bvb_{\so},\bub_{\so}) \;Z_{a_{\st},b_{\so}}(\buc_{\st};\bub_{\st}|\bvc_{\so};\bvb_{\so})
 Z_{a_{\so},b_{\st}}(\bub_{\so};\buc_{\so}|\bvb_{\st};\bvc_{\st}).
 \end{multline}
{\sl Remark 2.} We stress that \eqref{Resh-SP-eig1} is an identity. It is clear that it holds if
$\{\buc,\bvc\}\ne\{\bub,\bvb\}$, because in this case the r.h.s. of \eqref{Resh-SP-eig1} is the scalar product of two different on-shell vectors.
However, even in the case $\buc=\bub$ and $\bvc=\bvb$ the equation \eqref{Resh-SP-eig1} is still valid. This is because
the equation \eqref{Resh-SP-eig1} {\it does not describe} the norm of the on-shell vector at  $\buc=\bub$ and $\bvc=\bvb$.
In order to obtain the norm we first had to take the limit $\buc=\bub$ and $\bvc=\bvb$ in \eqref{Resh-SP} and only then use Bethe
equations. The way that we have used was opposite, therefore the r.h.s. of \eqref{Resh-SP-eig1} is identically zero for $a+b>0$.

The general formula \eqref{Resh-SP} also can be applied for the scalar products of the partial Bethe vectors
$ \mathbb{C}^{(l)}_{a,b}(\buc;\bvc)\mathbb{B}^{(l)}_{a,b}(\bub;\bvb)$, $l=1,2$. Then one should simply replace the functions
$r_k$ with the functions $r^{(l)}_k$.

\subsection{Calculating the sum over partitions}

We begin with equation \eqref{Mat-elS}.
We should substitute  the formulas for the scalar products of generic Bethe vectors \eqref{Resh-SP}
into the r.h.s. of this equation. Recall that we denote $r_k^{(1)}=\ell_k$, and thus,
$r_k^{(2)}=r_k\ell^{-1}_k$.

It is clear that in the end each set of the Bethe parameters  will be divided into four subsets. To avoid the cumbersome roman numbers,
we use arabic subscripts to denote these subsets. Namely, we assume that
\be{divis}
\begin{aligned}
\bucb_{\so}&\Rightarrow\{\bucb_1,\bucb_3\},\qquad \bucb_{\st}&\Rightarrow\{\bucb_2,\bucb_4\},\\
\bvcb_{\so}&\Rightarrow\{\bvcb_1,\bvcb_3\},\qquad \bvcb_{\st}&\Rightarrow\{\bvcb_2,\bvcb_4\}.
\end{aligned}
\ee
The cardinalities of the subsubsets are $a_n=\#\bucb_n$, $b_n=\#\bvcb_n$, $n=1,2,3,4$. In particular, $a_{\so}=a_1+a_3$
and $b_{\so}=b_1+b_3$.

We have
\begin{multline}\label{Mat-el1S}
\mathcal{M}^{(\kappa)}\left(m\Bigr|\begin{smallmatrix}
\buc & \bub \\
\bvc & \bvb
\end{smallmatrix}\right)^{a}_{b}=\sum
e^{\mathcal{Q}_{\bar\beta}+(\beta_2-\beta_1)a_{\so}}
\frac{\ell_1(\buc_{\st})\ell_3(\bvb_{\st})}{\ell_1(\bub_{\so})\ell_3(\bvc_{\so})}\;\ell_1(\bub_1)\ell_1(\buc_3)\ell_3(\bvb_1)\ell_3(\bvc_3)
\num
\times \; f(\buc_{\so},\buc_{\st})f(\bub_{\so},\bub_{\st})g(\bvc_{\st},\bvc_{\so}) g(\bvb_{\so},\bvb_{\st})
  f(\buc_1,\buc_3)  f(\bub_3,\bub_1)g(\bvc_3,\bvc_1)   g(\bvb_1,\bvb_3) f(\bvc_1,\buc_1)f(\bvb_3,\bub_3)\\
 \times
  \;Z_{a_3,b_1}(\buc_3;\bub_3|\bvc_1;\bvb_1) Z_{a_1,b_3}(\bub_1;\buc_1|\bvb_3;\bvc_3)\num
 \times\frac{r_1(\bub_2)r_1(\buc_4)r_3(\bvb_2)r_3(\bvc_4)}{\ell_1(\bub_2)\ell_1(\buc_4)\ell_3(\bvb_2)\ell_3(\bvc_4)}
  f(\buc_2,\buc_4)  f(\bub_4,\bub_2) g(\bvc_4,\bvc_2)   g(\bvb_2,\bvb_4) \frac{f(\bvc_2,\buc_2)f(\bvb_4,\bub_4)}
  {f(\bvc_{\st},\buc_{\st})f(\bvb_{\st},\bub_{\st})}\\
 \times
  \;Z_{a_4,b_2}(\buc_4;\bub_4|\bvc_2;\bvb_2)
 Z_{a_2,b_4}(\bub_2;\buc_2|\bvb_4;\bvc_4).
\end{multline}

Now we should express the products of the functions $r_k$ in \eqref{Mat-el1S} via (twisted) Bethe equations.
We have
\be{ru2}
 r_1(\bub_2)=\frac{f(\bub_2,\bub_1)f(\bub_2,\bub_3)f(\bub_2,\bub_4)}
 {f(\bub_1,\bub_2)f(\bub_3,\bub_2)f(\bub_4,\bub_2)}\;f(\bvb,\bub_2),
 \ee
\be{ru4}
 r_1(\buc_4)=e^{a_4(\beta_2-\beta_1)}\frac{f(\buc_4,\buc_1)f(\buc_4,\buc_2)f(\buc_4,\buc_3)}
 {f(\buc_1,\buc_4)f(\buc_2,\buc_4)f(\buc_3,\buc_4)}
 f(\bvc,\buc_4),
\ee
\be{rv2}
r_3(\bvb_2)=f(\bvb_2,\bub),
\ee
\be{rv4}
r_3(\bvc_4)=e^{b_4(\beta_2-\beta_3)}f(\bvc_4,\buc).
\ee
All these expressions should be substituted into \eqref{Mat-el1S}. We also should write the products over subsets
$\so$ and $\st$  in terms of the products over subsubsets \eqref{divis}. Then we obtain
\begin{multline}\label{Mat-el2}
\mathcal{M}^{(\kappa)}\left(m\Bigr|\begin{smallmatrix}
\buc & \bub \\
\bvc & \bvb
\end{smallmatrix}\right)^{a}_{b}
=\sum e^{\mathcal{Q}_{\bar\beta}+(\beta_2-\beta_1) (a-a_2)+(\beta_2-\beta_3)b_4}
\\
\times \frac{\ell_1(\buc_{2})\ell_1(\buc_{3})\ell_3(\bvb_{1})\ell_3(\bvb_{4})}{\ell_1(\bub_{2})\ell_1(\bub_{3})\ell_3(\bvc_{1})\ell_3(\bvc_{4})}
\; F^C_{uu}\;F^C_{vv}\;F^C_{vu}
\;F^B_{uu}\;F^B_{vv}\;F^B_{vu}\;  \mathcal{Z}.
\end{multline}
Here
\be{Z}
\mathcal{Z}=Z_{a_3,b_1}(\buc_3;\bub_3|\bvc_1;\bvb_1) Z_{a_1,b_3}(\bub_1;\buc_1|\bvb_3;\bvc_3)\num
Z_{a_4,b_2}(\buc_4;\bub_4|\bvc_2;\bvb_2) Z_{a_2,b_4}(\bub_2;\buc_2|\bvb_4;\bvc_4),
\ee
\be{FCuu}
F^C_{uu}=f(\buc_4,\buc_1)
f(\buc_3,\buc_2)
f(\buc_4,\buc_2)
f(\buc_4,\buc_3)
f(\buc_1,\buc_2)
f(\buc_1,\buc_3),
\ee
\be{FBuu}
F^B_{uu}=f(\bub_1,\bub_4)
f(\bub_2,\bub_3)
f(\bub_2,\bub_1)
f(\bub_2,\bub_4)
f(\bub_3,\bub_1)
f(\bub_3,\bub_4),
\ee
\be{FCvv}
F^C_{vv}=g(\bvc_4,\bvc_1)
g(\bvc_2,\bvc_3)
g(\bvc_2,\bvc_1)
g(\bvc_4,\bvc_2)
g(\bvc_3,\bvc_1)
g(\bvc_4,\bvc_3),
\ee
\be{FBvv}
F^B_{vv}= g(\bvb_1,\bvb_4)
g(\bvb_3,\bvb_2)
g(\bvb_1,\bvb_3)
g(\bvb_3,\bvb_4)
g(\bvb_1,\bvb_2)
g(\bvb_2,\bvb_4),
\ee
\be{FCvu}
F^C_{vu}=f(\bvc_1,\buc_4)f(\bvc_4,\buc_4)f(\bvc_1,\buc_1)f(\bvc_4,\buc_1)
f(\bvc_3,\buc_4)f(\bvc_4,\buc_3),
\ee
\be{FBvu}
F^B_{vu}=f(\bvb_3,\bub_3)f(\bvb_2,\bub_2)f(\bvb_3,\bub_2)f(\bvb_2,\bub_3)
f(\bvb_1,\bub_2)f(\bvb_2,\bub_1).
\ee

Actually it remains to combine different factors in \eqref{Mat-el2} together.
First of all we combine subsubsets into new groups:
\be{comb}
\begin{aligned}
\{\bucb_1,\bucb_4\}&=\bucb_{\rm i},\qquad \{\bucb_2,\bucb_3\}&=\bucb_{\rm ii},\\
\{\bvcb_1,\bvcb_4\}&=\bvcb_{\rm i},\qquad \{\bvcb_2,\bvcb_3\}&=\bvcb_{\rm ii}.
\end{aligned}
\ee
Then
\be{tFCuu}
F^C_{uu}=f(\buc_4,\buc_1)f(\buc_3,\buc_2)f(\buc_{\rm i},\buc_{\rm ii}),
\ee
\be{tFBuu}
F^B_{uu}=f(\bub_1,\bub_4)f(\bub_2,\bub_3)f(\bub_{\rm ii},\bub_{\rm i}),
\ee
\be{tFCvv}
F^C_{vv}= (-1)^{b_1b_{\rm ii} }g(\bvc_4,\bvc_1)g(\bvc_2,\bvc_3)g(\bvc_{\rm i},\bvc_{\rm ii}),
\ee
\be{tFBvv}
F^B_{vv}=(-1)^{b_1b_{\rm ii} }g(\bvb_1,\bvb_4)g(\bvb_3,\bvb_2)g(\bvb_{\rm ii},\bvb_{\rm i}),
\ee
\be{tFCvu}
F^C_{vu}=f(\bvc_4,\buc_3)f(\bvc_3,\buc_4)f(\bvc_{\rm i},\buc_{\rm i}),
\ee
\be{tFBvu}
F^B_{vu}=f(\bvb_1,\bub_2)f(\bvb_2,\bub_1)f(\bvb_{\rm ii},\bub_{\rm ii}).
\ee
Equation \eqref{Mat-el2} then takes the form
\begin{multline}\label{Mat-el4}
\mathcal{M}^{(\kappa)}\left(m\Bigr|\begin{smallmatrix}
\buc & \bub \\
\bvc & \bvb
\end{smallmatrix}\right)^{a}_{b}
=\sum_{\substack{\bucb\Rightarrow\{\bucb_{\rm i},\bucb_{\rm ii}\}\\
\bvcb\Rightarrow\{\bvcb_{\rm i},\bvcb_{\rm ii}\}}}
\frac{\ell_1(\buc_{\rm ii})\ell_3(\bvb_{\rm i})}{\ell_1(\bub_{\rm ii})\ell_3(\bvc_{\rm i})}\;G_1(\bucb_{\rm i};\bvcb_{\rm ii})
\, G_2(\bucb_{\rm ii};\bvcb_{\rm i})\\
\times f(\buc_{\rm i},\buc_{\rm ii})f(\bub_{\rm ii},\bub_{\rm i})g(\bvc_{\rm ii},\bvc_{\rm i})g(\bvb_{\rm i},\bvb_{\rm ii})
f(\bvc_{\rm i},\buc_{\rm i})f(\bvb_{\rm ii},\bub_{\rm ii}),
\end{multline}
where the functions $G_1$ and $G_2$ in their turn are given as sums over partitions:
 \begin{multline}\label{G1}
 G_1(\bucb_{\rm i};\bvcb_{\rm ii})=\sum_{\substack{\bucb_{\rm i}\Rightarrow\{\bucb_{1},\bucb_{4}\}\\
\bvcb_{\rm ii}\Rightarrow\{\bvcb_{2},\bvcb_{3}\}}}   f(\buc_{4},\buc_{1})  f(\bub_{1},\bub_{4})  g(\bvc_{2},\bvc_{3})
   g(\bvb_{3},\bvb_{2})
  \num
 \times f(\bvc_{3},\buc_{4})f(\bvb_{2},\bub_{1}) \;Z_{a_{4},b_{2}}(\buc_{4};\bub_{4}|\bvc_{2};\bvb_{2})
 Z_{a_{1},b_{3}}(\bub_{1};\buc_{1}|\bvb_{3};\bvc_{3}),
 \end{multline}
and
 \begin{multline}\label{G2}
 G_2(\bucb_{\rm ii};\bvcb_{\rm i})=\sum_{\substack{\bucb_{\rm ii}\Rightarrow\{\bucb_{2},\bucb_{3}\}\\
\bvcb_{\rm i}\Rightarrow\{\bvcb_{1},\bvcb_{4}\}}}  e^{\mathcal{Q}_{\bar\beta}+(\beta_2-\beta_1) (a-a_2)+(\beta_2-\beta_3)b_4}
  f(\buc_{3},\buc_{2})f(\bub_{2},\bub_{3})
  \num
 \times  g(\bvc_{1},\bvc_{4})
   g(\bvb_{4},\bvb_{1})f(\bvc_{4},\buc_{3})f(\bvb_{1},\bub_{2}) \;Z_{a_{3},b_{1}}(\buc_{3};\bub_{3}|\bvc_{1};\bvb_{1})
 Z_{a_{2},b_{4}}(\bub_{2};\buc_{2}|\bvb_{4};\bvc_{4}).
 \end{multline}
Comparing the function $G_1$ \eqref{G1} with \eqref{Resh-SP-eig1} we see that $G_1=0$ unless
$\bucb_{\rm i}=\emptyset$ and $\bvcb_{\rm ii}=\emptyset$. Hence, $\bucb_{\rm ii}=\bucb$,
$\bvcb_{\rm i}=\bvcb$, $a_1=a_4=0$, $b_2=b_3=0$.

Looking now at $G_2$ \eqref{G2} and comparing it with \eqref{Resh-SP-eig} we see that
\be{G2-S}
G_2= e^{\mathcal{Q}_{\bar\beta}}\mathcal{S}^{(\kappa)}_{a,b}.
\ee
Substituting this into \eqref{Mat-el4} and setting there $\bucb_{\rm i}=\emptyset$ and $\bvcb_{\rm ii}=\emptyset$ we immediately
arrive at
\begin{equation}\label{Mat-el6}
\mathcal{M}^{(\kappa)}\left(m\Bigr|\begin{smallmatrix}
\buc & \bub \\
\bvc & \bvb
\end{smallmatrix}\right)^{a}_{b}
=e^{\mathcal{Q}_{\bar\beta}}\frac{\ell_1(\buc)\ell_3(\bvb)}{\ell_1(\bub)\ell_3(\bvc)}\;\mathcal{S}^{(\kappa)}_{a,b},
\end{equation}
and thus, proposition~\ref{thmM} is proved.


\begin{thebibliography}{99}
\bibitem{TakF79}
L. A. Takhtadzhan and L. D. Faddeev, {\sl The quantum method of the inverse problem and the
Heisenberg XY Z model}, Russian~Math.~Surveys {\bf 34}:5 (1979), 11--68.
%
\bibitem{SklTF79}
E. K. Sklyanin,
L. A. Takhtadzhyan and L. D. Faddeev, {\sl Quantum inverse problem method. I},
Theoret. and Math. Phys. {\bf 40}:2 (1979), 688--706.
%
\bibitem{FadLH96} L. D. Faddeev, in: Les Houches Lectures {\sl Quantum Symmetries}, eds A. Connes
et al, North Holland, (1998) 149.
%
\bibitem{KorBI93L}
V. E. Korepin, N. M. Bogoliubov, and A. G. Izergin, {\sl Quantum inverse scattering method and correlation functions},
Cambridge Monogr. Math. Phys., Cambridge Univ. Press, Cambridge 1993.
%
\bibitem{IzeK84} A. G. Izergin,  V. E. Korepin,
\textsl{The quantum inverse scattering method approach to correlation functions},
Comm. Math. Phys. {\bf 94} (1984) 67--92.
%
\bibitem{KitMST05} N. Kitanine, J. M. Maillet, N. A. Slavnov and V. Terras,
\textsl{Master equation for spin-spin correlation functions of the
$XXZ$ chain}, Nucl. Phys. B {\bf 712} (2005) 600--622, \texttt{arXiv:hep-th/0406190}.
%
\bibitem{Sla07} N. A. Slavnov, {\sl The algebraic Bethe ansatz and quantum integrable systems}, Russian Math. Surveys, {\bf 62}:4 (2007), 727--766.
%
\bibitem{PakRS15d} S. Pakuliak, E. Ragoucy, N. A. Slavnov, \textsl{$GL(3)$-Based Quantum Integrable Composite Models. I.
Bethe vectors}, SIGMA {\bf 11} (2015) 063, 20 pp., \texttt{arXiv:1501.07566}.
%
\bibitem{PakRS15b} S. Pakuliak, E. Ragoucy, N. A. Slavnov, \textsl{$GL(3)$-Based Quantum Integrable Composite Models. II.
Form Factors of Local Operators}, SIGMA {\bf 11} (2015) 064, 18 pp., \texttt{arXiv:1502.01966}.
%
\bibitem{PakRS15a} S. Pakuliak, E. Ragoucy, N. A. Slavnov, \textsl{Zero modes method and form factors in quantum integrable models},
Nucl. Phys. B {\bf 893} (2015) 459--481, \texttt{arXiv:1412.6037}.
%
\bibitem{KitMT99} N. Kitanine, J. M. Maillet and V. Terras, \textsl{Form factors of the XXZ Heisenberg spin-12 finite chain},
Nucl. Phys. B  {\bf 554}  (1999) 647--678, \texttt{arXiv:math-ph/9807020}.
%
\bibitem{MaiT00} J. M. Maillet, V. Terras, \textsl{On the quantum inverse scattering problem}, Nucl. Phys. B
{\bf 575} (2000) 627--644, \texttt{hep-th/9911030}.
%
\bibitem{KitKMST11} N. Kitanine, K. Kozlowski, J. M. Maillet, N. A. Slavnov, V. Terras, {\sl
A form factor approach to the asymptotic behavior of correlation functions}, J. Stat. Mech. (2011) P12010,
\texttt{arXiv:hep-th/1110.0803}.
%
\bibitem{KitKMST12} N. Kitanine, K. Kozlowski, J. M. Maillet, N. A. Slavnov, V. Terras, {\sl
Form factor approach to dynamical correlation functions in critical models}, J. Stat. Mech. (2012) P09001,
\texttt{arXiv:1206.2630}.
%
\bibitem{CauM05}
J.~S. Caux and J.~M. Maillet, {\sl Computation of Dynamical Correlation Functions of
Heisenberg Chains in a Magnetic Field},
Phys. Rev. Lett. {\bf 95} (2005) 077201 3pp, \texttt{arXiv:cond-mat/0502365}.
%
\bibitem{CauPS07}J.~S. Caux, P. Calabrese, N.~A. Slavnov, {\sl One-particle dynamical correlations in the
one-dimensional Bose gas},  J. Stat. Mech. (2007) P01008, \texttt{arXiv:cond-mat/0611321}.
%
\bibitem{HutLPRS16a} A. Hutsalyuk, A. Liashyk, S. Z. Pakuliak, E. Ragoucy, N. A. Slavnov,
\textsl{Form factors of the monodromy matrix entries in $\mathfrak{gl}(2|1)$-invariant integrable models},
Nucl. Phys. B {\bf 911} (2016) 902--937, \texttt{arXiv:1607.04978}.
%
%
\bibitem{Sch87} P. Schlottmann, \textsl{Integrable narrow-band model with possible relevance to heavy
Fermion systems}, Phys. Rev. B {\bf 36} (1987) 5177--5185.
%
\bibitem{For89} D. F\"orster, \textsl{Staggered spin and statistics in the supersymmetric t-J model}, Phys.
Rev. Lett. {\bf 63} (1989) 2140--2143.
%
\bibitem{EssK92} F. H. L. Essler and V. E. Korepin, \textsl{Higher conservation laws and algebraic Bethe
Ansatze for the supersymmetric t-J model}, Phys. Rev. B {\bf 46} (1992) 9147--9162.
%
\bibitem{FoeK93} A. Foerster and M. Karowski, \textsl{Algebraic properties of the Bethe ansatz for an
$spl(2,1)$-supersymmetric t-J model}, Nucl. Phys. B {\bf 396} (1993) 611--638.
%
\bibitem{Gom02} F. G\"ohmann, \textsl{Algebraic Bethe ansatz for the $gl(1|2)$ generalized model and Lieb--Wu
equations}, Nucl. Phys. B {\bf 620} (2002) 501--518, \texttt{arXiv:cond-mat/0108486}.
%
\bibitem{GohK00} F. G\"ohmann, V. E. Korepin, \textsl{Solution of the quantum inverse problem},
J. Phys. A: Math. Gen., {\bf 33} 2000 1199--1220, \texttt{arXiv:hep-th/9910253}.
%
\bibitem{PakRS15c} S. Pakuliak, E. Ragoucy, N. A. Slavnov, \textsl{Form factors of local operators in a one-dimensional two-component Bose gas},
J. Phys. A, {\bf 48}:43 (2015) 435001, 21 pp., \texttt{arXiv:1503.00546}.
%
\bibitem{PakRS17} S. Z. Pakuliak, E. Ragoucy, N. A. Slavnov, \textsl{Bethe vectors for models based on the super-Yangian $Y(\mathfrak{gl}(m|n))$}, \texttt{arXiv:1604.02311}.
%
\bibitem{KulS80} P. P. Kulish and E. K. Sklyanin, \textsl{On the solution of the Yang-Baxter equation},
Zap. Nauchn. Semin. LOMI {\bf 95} (1980) 129--160;  J. Sov. Math. {\bf 19} (1982) 1596 (Engl. transl.).
%
\bibitem{BeRa08}
S. Belliard and E. Ragoucy, {\sl The nested Bethe ansatz for 'all' closed spin chains.},
 J. Phys. A {\bf 41} (2008) 295202, \texttt{arXiv:0902.0321}.
%
\bibitem{HutLPRS17a} A.~A.~Hutsalyuk,  A.~N.~Liashyk, S.~Z.~Pakuliak, E.~Ragoucy, N.~A.~Slavnov,
\textsl{Current presentation for the double super-Yangian $DY(\mathfrak{gl}(m|n))$ and Bethe vectors},
Russ. Math. Surv. (2017) (in print), \texttt{arXiv:1606.03573}.
%
\bibitem{Fuk17} J.~Fuksa, \textsl{Bethe vectors for composite generalised models
with $\mathfrak{gl}(2|1)$ and $\mathfrak{gl}(1|2)$ supersymmetry}, (2016), \texttt{arXiv:1611.00943}.
%
%
\bibitem{HutLPRS16b} A. Hutsalyuk, A. Liashyk, S. Z. Pakuliak, E. Ragoucy, N. A. Slavnov,
\textsl{Scalar products of Bethe vectors in models with  $\mathfrak{gl}(2|1)$ symmetry 2. Determinant representation},
J. Phys. A: Math. Theor., {\bf 50}:3 (2017), 034004,   \texttt{arXiv:1606.03573}.
%
\bibitem{HutLPRS16c} A. Hutsalyuk, A. Liashyk, S. Z. Pakuliak, E. Ragoucy, N. A. Slavnov,
\textsl{Scalar products of Bethe vectors in models with  $\mathfrak{gl}(2|1)$ symmetry 1. Super-analog of Reshetikhin formula},
J. Phys. A: Math. Theor., {\bf 49}:45 (2016), 454005,   \texttt{arXiv:1605.09189}.
%
\bibitem{BraGLZ95} A. J. Bracken, M. D. Gould, J. R. Links, Y. Z. Zhang, \textsl{New Supersymmetric
and Exactly Solvable Model of Correlated Electrons}, Phys. Rev. Lett. {\bf 74}:14 (1995), 2768--2771,
\texttt{arXiv:cond-mat/9410026}.
%
\bibitem{BedF95} G. Bed\"urftig, H. Frahm \textsl{Thermodynamics of an integrable model
for electrons with correlated hopping},
J. Phys. A: Math. Gen., {\bf28} (1995) 4453,  \texttt{arXiv:cond-mat/9504103}.
%
\bibitem{RamM96} P. B. Ramos, M. J. Martins, \textsl{One parameter family of an integrable $spl(2|1)$ vertex model:
Algebraic Bethe ansatz approach and ground state structure},
Nucl. Phys. B {\bf474} (1996) 678--714, \texttt{arXiv:hep-th/9604072}.
%
\bibitem{PfaF96} M. P. Pfannm\"uller, H. Frahm
\textsl{Algebraic Bethe Ansatz for $gl(2,1)$ Invariant 36-Vertex Models},
Nucl. Phys. B {\bf 479} (1996) 575--593, \texttt{arXiv:cond-mat/9604082}.

\end{thebibliography}
\end{document}